\documentclass{article}

\usepackage{microtype}
\usepackage{graphicx}
\usepackage{subfigure}
\usepackage{booktabs} % for professional tables
\usepackage{multirow}
\usepackage{hyperref}

\usepackage{hyperref}

\usepackage[accepted]{icml2021}

\icmltitlerunning{Personalized and Reliable Decision Sets}

\begin{document}

\twocolumn[
\icmltitle{Personalized and Reliable Decision Sets: \\
           Enhancing Interpretability in Clinical Decision Support Systems}



\icmlsetsymbol{equal}{*}

\begin{icmlauthorlist}
\icmlauthor{Francisco Valente}{cisuc}
\icmlauthor{Simão Paredes}{isec,cisuc}
\icmlauthor{Jorge Henriques}{cisuc}
\end{icmlauthorlist}

\icmlaffiliation{cisuc}{Centre for Informatics
and Systems of University of Coimbra, University of Coimbra, Portugal}

\icmlaffiliation{isec}{Coimbra
Institute of Engineering (ISEC), Polytechnic of Coimbra, Portugal}

\icmlcorrespondingauthor{Francisco Valente}{paulo.francisco.valente@gmail.com}

\icmlkeywords{}

\vskip 0.3in]



\printAffiliationsAndNotice{}  % leave blank if no need to mention equal contribution

\begin{abstract}
In this study, we present a novel clinical decision support system and discuss its interpretability-related properties. It combines a decision set of rules with a machine learning scheme to offer global and local interpretability. More specifically, machine learning is used to predict the likelihood of each of those rules to be correct for a particular patient, which may also contribute to better predictive performances. Moreover, the reliability analysis of individual predictions is also addressed, contributing to further personalized interpretability. The combination of these several elements may be crucial to obtain the clinical stakeholders' trust, leading to a better assessment of patients' conditions and improvement of the physicians' decision-making.
\end{abstract}

\section{Contextualization}
\label{context}

With the advent of machine learning (ML) in healthcare, the urge for the development of interpretable methods has grown. Even there is a lack of agreement on the definition of interpretability, it is commonly associated with the transparency of the several elements (features, algorithms, parameters, generated model, etc.) of a decision system \cite{Ahmad2018}. Interpretability of ML models is particularly important in healthcare, considering multiple perspectives, such as the medical, patient, and legal ones \cite{Amann2020}. It ensures the understanding of the reasoning behind the decisions or recommendations of those models and bias identification \cite{Ahmad2018}, increasing the trust of their users. In short, interpretability is a decisive element to the broader adoption of ML-based methodologies in healthcare.

Several categorizations of interpretability models have been proposed, being (1) global or local, and (2) ante-hoc or post-hoc, two of the most frequently considered classifications \cite{Stiglic2020}. Some researchers have advocated the preference of intrinsically interpretable models (ante-hoc) in high-stakes domains like healthcare, over models that extract explanations from “black-box” approaches (post-hoc), as those explanations do not truly reveal the reasoning of such models \cite{Rudin2019, Laugel2019}.

Both global and local interpretability offer valuable contributions from a healthcare perspective. On the one hand, global methods imply an overall understanding of the relationship between the inputs and outputs of a model to the entire study population \cite{Ahmad2018, Stiglic2020}, which is imperative to obtain the trust of physicians. On the other hand, local models infer explanations at the individual level, which allows the user to better understand the decisions/recommendations behind each instance. The latter leads the clinicians to further assess and forecast the patient’s condition, and thus improve their decisions (diagnosis, treatment, etc) – personalized interpretability.

However, personalization can also be employed in terms of the ML itself. Such characteristic is often applied in recommendation systems or web personalized search, which considers individual-specific information to produce tailored predictions \cite{Schneider2019}. However, often, ML is used to produce only a general model, which is optimized to performs the best for the population level. In most healthcare scenarios, there is no previous patient-specific information to be used to personalize a model. Even so, one may explore strategies to create systems that adjust a global model to individual instances. Considering then personalized decision models, it is expected to obtain improved diagnosis/forecasting of patient conditions. 

Besides such attributes, the assessment of the model's reliability is also considered by the physicians as a kind of explanation that supports the output \cite{Tonekaboni2019}. More specifically, an estimation (e.g., a score) that may inform the physicians if the output obtained with the decision model is reliable, and then the clinicians can trust it, or if it is likely to be incorrect, and thus a more cautious analysis is required before the algorithm’s predictions or recommendations are accepted. Thus, the developed clinical decision support systems should not only provide explanations of their decisions but also provide awareness of when they may not be reliable \cite{Tonekaboni2019}. 

Even a high-performing model will be incorrect for some instances. In high-stakes domains like healthcare, knowing whose predictions are likely to be misleading is crucial, as an individual error may result in critical outcomes. Although some research has been performed about this assessment \cite{Bosnic2008a, Jiang2018, Schulam2019}, its study remains largely unexplored, despite its undeniable importance. In fact, it has been revealed that physicians have overconfidence in their diagnostics/decisions, and that such cognitive bias is a major cause of medical decision-making errors \cite{Berner2008, Croskerry2008}. While modeling the confidence level of physicians seems to be implausible, to predict the trustworthy degree of algorithms appears to be not only possible as highly desirable. To be able to inform when the algorithms’ outputs are unreliable may be a peremptory characteristic to foster the physicians’ trust. 

Several methodologies that have been proposed to promote interpretability in ML models are rule-based ones, such as falling rule lists \cite{Wang2015}, interpretable decision sets \cite{Lakkaraju2016}, MUSE \cite{Lakkaraju2018}. In fact, decision rules are widely applied in clinical practice to support and guide physicians’ decisions, enhancing their evidence-based extent and reducing variation among clinicians \cite{Brehaut2005}. Furthermore, decision rules are broadly acknowledged as one of the most simple, intelligible, and useful methods to describe cause-effect relationships and to represent domain knowledge, two major concerns in healthcare decision-making. Therefore, incorporating decision rules as the main element in interpretability-based methodology seems a logical choice.

\section{Personalized and Reliable Decision Sets}

\begin{figure*}[h!!]
\vskip 0.2in
\begin{center}
\centerline{\includegraphics[width=\textwidth]{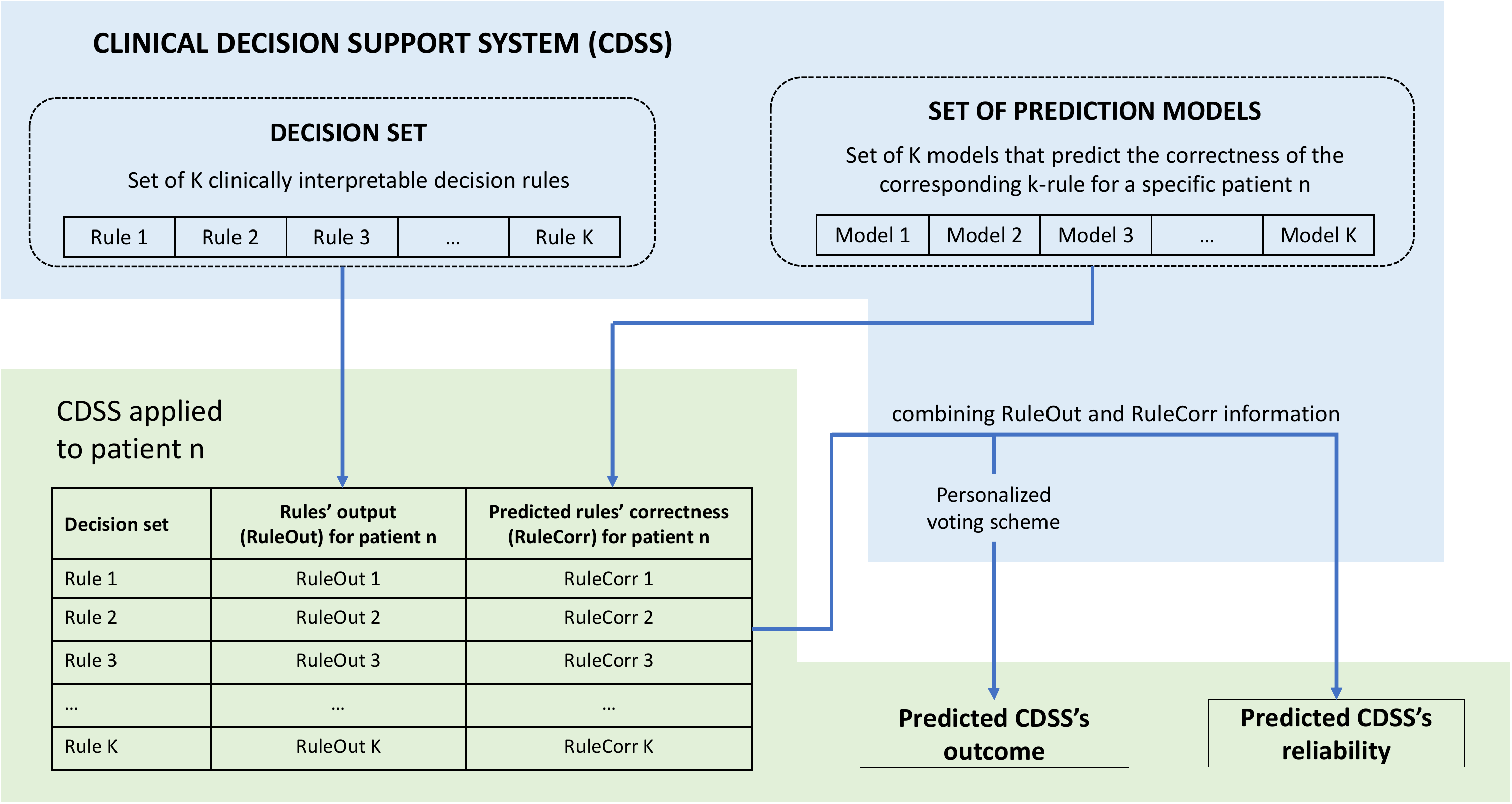}}
\caption{General scheme of the clinical decision support system and its application.}
\label{fig:cdss_scheme}
\end{center}
\vskip -0.2in
\end{figure*}

We intend to provide an analysis of a novel clinical decision support system (CDSS). The related methodology has been applied to healthcare scenarios, namely events prediction after acute coronary syndromes \cite{Valente2021} and diseases' diagnosis \cite{valente2021improving}. In this paper, that CDSS will be presented considering a broader perspective, namely its application to any binary problem that uses tabular data. We will also further discuss how its several elements provide important contributions to model interpretability in the healthcare domain. Figure \ref{fig:cdss_scheme} shows an overall representation of the studied CDSS. Its several parts are described and discussed in sections \ref{subsec:decision set}-\ref{subsec:reliability}, and some practical examples and results are presented to support the analysis.

\subsection{Decision Set of Clinically Interpretable Rules}
\label{subsec:decision set}

The methodology is based on the use of decision rules. More specifically, a decision set. A decision set is a collection of rules, where each rule is applied individually and without following any specific order. These sets are considered to have some more key interpretability characteristics than other rule-based methods like decision trees or decisions lists \cite{Lakkaraju2016}. Considering the proposed aim, any type of decision rules may be employed: data-driven rules, knowledge-driven ones, or a combination of both.

Knowledge-driven rules are created based on the clinical literature or the clinical experience of medical practitioners. Thus, they may facilitate the physicians’ trust. However, sometimes those rules may not be available, which implies that data-driven rules must be generated. Furthermore, an embedded data-driven rules procedure may be a useful tool as patients’ characteristics may differ across populations, and so the rules created for a given population may not be fully appropriate for a different population. Thus, data-driven rules are expected to be adjusted to the population where the model is applied. Several methodologies have been proposed to generate rules by discovering patterns from data \cite{Lakkaraju2016}.

However, these data-driven rules must be validated by medical experts. In fact, clinicians expect that the model’s features and represented knowledge are aligned with the evidence-based medical practice \cite{Tonekaboni2019}. Such issues are requisites for global interpretability and a critical element to obtaining the physicians’ trust in the generated model. Hybrid approaches, which take into account domain knowledge, are a possible procedure to ensure that the created data-driven rules might be applied in clinical practice. For example, to consider only features whose scenario-specific relevance has been proven in large clinical studies \cite{Valente2021}.

We propose that the decision set in this CDSS must preferably be composed of two-way decision rules: \textit{IF (rule's condition), THEN (rule's output 1), ELSE (rule's output 2)}. Theoretically, the CDSS can also considers a set of one-way decision rules: \textit{IF (rule's condition), THEN (rule's output)}. However, it has some practical limitations that will be discussed in section \ref{subsec:personalization}. Even one-way decision sets are most common in the literature, some methods have been proposed to develop sets of data-driven two-way decision rules, for classification \cite{Benard2021a, Valente2021, valente2021improving} and regression \cite{Benard2021} scenarios. An example of a two-way rules' set will be later presented in Table \ref{tab:case_study}.
 
Finally, two elements must be constrained in the decision set: the number of decision rules and the length of those rules. Intuitively, the smaller the number of rules in the set and the shorter the rules, the higher the interpretability.

\subsection{Personalization of the Decision Set using Machine Learning}
\label{subsec:personalization}

Often, the outcome produced by a decision set results from combining the votes of the individual decision rules’ prediction, using averaging or majority voting mechanisms. Those voting schemes may also be weighted, usually based on the individual rules' accuracy on the training population. In this CDSS, the weighted voting is considered from a personalized perspective. More specifically, the aim is to predict if each rule of the decision set will be (or not) correct for a given new patient, and then consider such predictions to weigh the rules' outputs for that patient.

For example, let us consider the two-way rule \textit{IF age$>$80 years AND comorbidities$>$1, THEN death, ELSE survival}. This rule describes a general tendency: the majority of patients with more than 80 years and 2 or more comorbidities will die. However, it does not apply to all patients with those characteristics. Similarly, not all patients with less than 80 years and less than 2 comorbidities will survive. If this rule was correct for 90\% of the training instances, then in a non-personalized weighted voting, that rule would have a weight of 0.9. However, in this novel methodology, the focus is not in the accuracy of the rule for the overall population but its accuracy for a particular patient.

The goal is then to forecast the patients where that rule will be correct and the ones where it will be not. This introduces personalization in the methodology as the model will specify how an individual decision rule is expected to behave for each patient: the same rule will have a different importance for each individual. To the best of our knowledge, no similar procedure (personalize a given decision set) was considered before. 

\begin{figure*}[h!!]
\vskip 0.2in
\begin{center}
\centerline{\includegraphics[width=\textwidth]{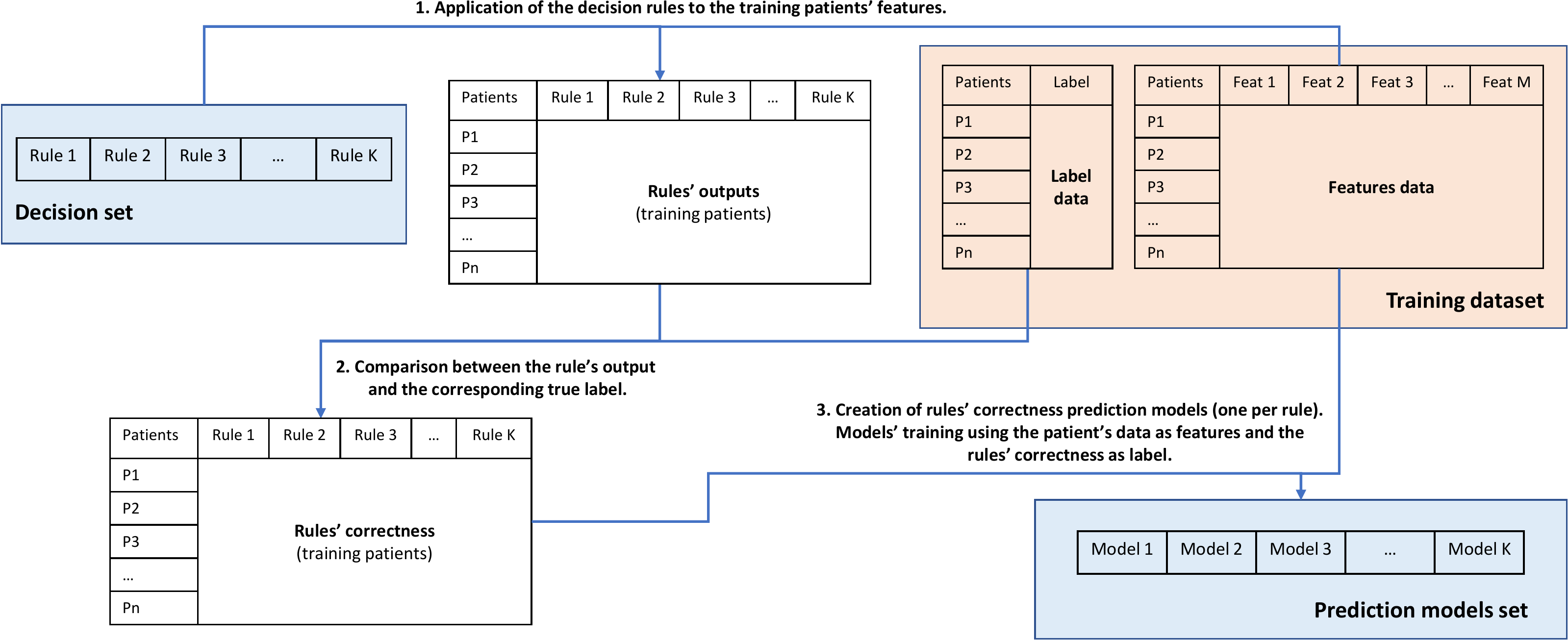}}
\caption{General scheme of the creation of the rule's correctness prediction models.}
\label{fig:models_scheme}
\end{center}
\vskip -0.2in
\end{figure*}

In order to produce that rule's correctness forecasting, it is required a prediction model. This CDSS proposes to achieve it by training a model using a machine learning (ML) process.  Figure \ref{fig:models_scheme} presents a general scheme about how the prediction models are generated in this methodology.

In a standard ML procedure, we predict the label of a given sample (e.g., the patient will die/survive, the patient’s cancer is benign/malign). In this methodology, we attempt to predict the correctness of a rule for a given sample, and thus that rule’s correctness will be our label. Even those labels are not available at first, they are easy to obtain for the training dataset. Intuitively, they are created comparing the outcome predicted by a rule and the true outcome (given by the known target). In other words, for each rule in the decision set, a Nx1 rule’s correctness vector is created, where N is the number of samples (patients) used to train the model - Figure \ref{fig:models_scheme}, step 2. This vector is then binary, with a value of 1 if the output predicted by the rule and the true outcome are the same, and a value of 0 otherwise. Table \ref{tab:rule_corr} exemplifies how those labels are obtained for a given two-way rule.

% TABLE 1
% \usepackage{multirow}
\begin{table*}[]
\centering
\caption{Example of how to obtain the rule's correctness label. The presented decision rule and the input-and-output data for patients are arbitrary and for demonstration purposes.}
\label{tab:rule_corr}
\begin{tabular}{lcccccc}
\hline
\multirow{2}{*}{Decision rule} &
  \multirow{2}{*}{Patients} &
  \multicolumn{2}{c}{Feature values} &
  \multirow{2}{*}{Rule's  output} &
  \multirow{2}{*}{True label} &
  \multirow{2}{*}{\begin{tabular}[c]{@{}c@{}}\textbf{New label:}\\  \textbf{Rule's correctness}\end{tabular}} \\ \cline{3-4}
 &     & age & nc  &     &     &     \\ \hline
\multirow{5}{*}{\begin{tabular}[c]{@{}l@{}}IF age\textgreater{}80 AND number \\ of comorbidities (nc)\textgreater{}1,\\ THEN death (1),\\ ELSE survival(0)\end{tabular}} &
  P1 &
  90 &
  3 &
  1 &
  1 &
  \textbf{1} \\
 & P2   & 47  & 1   & 0   & 1   & \textbf{0}   \\
 & P3   & 82  & 0   & 0   & 0   & \textbf{1}   \\
 & P4   & 86  & 2   & 1   & 0   & \textbf{0}   \\
 & ... & ... & ... & ... & ... & \textbf{...} \\ \hline
 
\end{tabular}
\end{table*}

Therefore, instead of a single label, there will be as many labels as rules, each one related to an individual rule’s correctness. In other words, we will have a predictor for each rule. Having the labels’ vectors, it is also required a set of features to train the prediction models. We propose to use the available independent variables' values, i.e., the data that would be used to predict the original label in a standard ML procedure or that is used to derive data-driven rules - Figure \ref{fig:models_scheme}, step 3.

Any classification algorithm may be applied to train/develop the prediction models. However, if a fully interpretable approach is required, a more easy-to-understand algorithm, such as a logistic regression method, is advisable (ante-hoc interpretability). 

As abovementioned, those prediction models will then be used to predict the correctness of a rule (i.e., its likelihood of being correct) for a new patient with an unknown target, and that predicted correctness value is used to weigh the rule's output for that particular patient. Overall, the idea is that the contribution of an individual decision rule to the final outcome is proportional to its predicted correctness. In other words, for a particular patient, rules with a higher likelihood of being correct (higher rule's correctness prediction value) will be given higher importance. For example, considering a personalized weighted average as the voting scheme of all the K rules in the decision set, the outcome for a patient $n$ may be defined as:

\begin{equation}
 Prob_{n} (class=1) = {\frac {\sum _{i=1}^{K}rule\_output_{i,n}\cdot weight_{i,n}}{\sum_{i=1}^{K}weight_{i,n}}},
\end{equation}

While in a non-personalized weighted average, the weight $i$ would be the same for any $n$ (the weight of a given rule is the same for all patients), is this personalized approach that weight is dependent on the patient $n$ as well. As this personalization leads to a patient-fitted formulation, it is theoretically expected to have an associated improvement in the prediction performance compared to the standard weighting. In any case, in an averaging voting, the result should be calibrated to produce a clinically meaningful probability \cite{VanCalster2019}. 

In order to inspect the predictive impact of the personalized procedure, we applied the CDDS to three public clinical datasets: Heart Disease (Heart), Breast Cancer Wisconsin Diagnostic (Breast), and Mammographic Mass (Mammo)\footnote{The three datasets are available at \url{https://archive.ics.uci.edu}.}. In the three scenarios, the goal is to predict the presence or absence of disease. The decision support system was defined as: (1) a decision set of 10 two-way short decision rules was created following the approach considered in  \cite{valente2021improving} - it combines extraction of rules from an ensemble of small-depth decision trees with LASSO subset selection to obtain a set with a constrained number and length of rules; (2) rule's correctness prediction models were obtained following the procedure described in section \ref{subsec:personalization}, considering logistic regression as the machine learning method; (3) the final outcome was computed through a personalized voting method suggested in \cite{Valente2021}. 

The area under the receiver operating characteristic curve (ROC AUC) results obtained for that decision system, using a 5-times repeated 5-fold cross-validation, are presented in Table \ref{tab:AUCS} (\textit{Personalized decision set}). In the same table, those results are compared to baseline decision set' outcome prediction methods, considering the same set of decision rules. More specifically, a simple averaging of the rules' outputs (\textit{Non-weighted decision set}), and a non-personalized weighted averaging of the rules' outputs (\textit{Weighted decision set}) where the weights are defined by the overall rules' accuracy instead of a patient-specific one.

\begin{table}[]
\centering
\caption{Area under the ROC curve (AUC) values for the three clinical datasets (Heart, Breast and Mammo), considering the proposed methodology (in bold) and baseline methods.}
\label{tab:AUCS}
\begin{tabular}{lccc}
\hline
Method                             & \multicolumn{3}{c}{AUC ROC}                         \\ \hline
                                   & Heart        & Breast       & Mammo        \\ \hline
Non-weighted decision set     & 0.82 & 0.98   & 0.84 \\
Weighted decision set          & 0.84  & 0.98   & 0.85 \\
\textbf{Personalized decision set} & \textbf{0.89} & \textbf{0.99}  & \textbf{0.90 } \\ \hline
\end{tabular}
\end{table}

The preliminary results of Table \ref{tab:AUCS} suggest that the personalization mechanism, incorporated in this decision system through the prediction of the rules' correctness for a given patient, can indeed provide a significant improvement of the predictive ability performance of the decision set, as theoretically expected.

The personalization of the decision set appears to be useful to improve the predictive performance, and it has also advantages in terms of individual-level interpretability. For better clarification, Table \ref{tab:case_study} presents a hypothetical scenario, with a set of clinical rules, and the predicted probability values for the correctness of each decision rule, for a particular patient whose characteristics are detailed in the caption of the table.

\begin{table*}[]
\centering
\caption{Example of the application of a decision set to a patient, and the predicted correctness values of the rules for that patient. Patient's characteristics: age = 86, male, number of comorbidities (nc) = 2,  with antecedents of stroke (ac), no smoker. The presented decision rules and the related predicted rule's correctness values are arbitrary and for demonstration purposes.}
\label{tab:case_study}
\begin{tabular}{lcc}
\hline
\textbf{Decision set} &
  \textbf{Rule's output} &
  \textbf{\begin{tabular}[c]{@{}c@{}}Predicted rule's\\ corretness\end{tabular}} \\ \hline
\begin{tabular}[c]{@{}l@{}}Rule 1: \\ IF age\textgreater{}80 AND nc\textgreater{}1, THEN death(1), ELSE survival(0)\end{tabular} &
  1 &
  66\% \\
\begin{tabular}[c]{@{}l@{}}Rule 2: \\ IF nc\textgreater{}3, THEN death(1), ELSE survival(0)\end{tabular}      & 0   & 42\% \\
\begin{tabular}[c]{@{}l@{}}Rule 3:\\ IF male=yes AND smoker=yes, THEN death(1), ELSE survival(0)\end{tabular} & 0   & 54\% \\
\begin{tabular}[c]{@{}l@{}}Rule 4:\\ IF ac=yes AND male=yes AND age\textgreater{}65, THEN death(1), ELSE survival(0)\end{tabular} &
  1 &
  95\% \\
...                                                                                                           & ... & ...  \\ \hline
\end{tabular}
\end{table*}

Table \ref{tab:case_study} simulates then the obtained estimations for a given instance. Decision rules with the same output (rules 1 and 4; rules 2 and 3) may significantly differ in the probability of being correct, based on what was verified in previous patients. Moreover, while the number of rules that predict the positive output (death) is the same as the number of negative ones (survival), the predicted correctness values suggest that the rules that predict the patient’s death are more likely to be correct. Thus, from the set of decision rules, the physicians may analyze the ones that are expected to contribute more to a correct forecasting. Consecutively, the clinicians may also verify which risk factors are more likely to be important. For example, in this scenario, rule 4 has a remarkable probability of being correct for this patient. Therefore, a (total or partial) combination of three risk factors – antecedents of stroke, male, 86 years old - seems to be very relevant for this individual.

In short, the methodology starts from a set of (clinically meaningful) decision rules, which describe the problem as a whole through a group of interpretable cause-effect relationships - global interpretability. However, the approach considers an additional procedure where it is evaluated the likelihood of each decision rule to be correct for a particular patient – local/personalized interpretability. 

The question of whether one-way decision rules may be applied or not in this approach is associated with the computation of the rule’s correctness label and the creation of the corresponding prediction models. One-way rules do not have an \textit{ELSE} segment, and then they do not apply to all instances - e.g., in Table \ref{tab:rule_corr}, the decision rule would not apply to patients 2 and 3 if we considered only the \textit{THEN} element. Thus, it is not possible to compute a rule's correctness value for such instances. Considering that, some issues arise. First, the rule's correctness label would be available for different patients for each rule, in the training phase of the prediction models. Second, a given rule could potentially apply only to a small number of instances, and thus the samples available to train the models would probably be too scarce to produce good prediction models. Third, a given rule could be correct (or incorrect) for all the (few) instances it applies, and thus the rule's correctness label would have only one class, and the creation of prediction models would not be possible. The latter consideration is implausible to occur in the two-way rules scenario. Therefore, while the use of one-way decision rules is theoretically possible,  some practical concerns require further analysis. However, the study of when such limitations impose and/or how to overcome them was not a focus in this paper.

\subsection{Reliability Estimation for Single Predictions}
\label{subsec:reliability}

As discussed in the Contextualization section, the reliability analysis may be a very relevant criterion to further evaluate the prediction of a decision model in healthcare, contributing to a higher level of interpretability. Even so, that assessment is rarely considered. In this decision support system, the reliability estimation is obtained using the information provided by the output and correctness values predicted for each rule. However, the way those values are used is different and independent of the voting mechanism that produces the final outcome in section \ref{subsec:personalization}.

For the reliability estimation, we may infer that in an ideal scenario, decision rules with the same output will have equivalent correctness. For example, if we are attempting to predict if a patient will die from a disease, the rules that predict a correct output will ideally be associated with a high correctness value, and the remaining rules will have a low correctness value. In other words, we are estimating that rules that produce opposite outputs have very different likelihoods of being correct. In that scenario, there is high confidence in the predicted output. However, if the correctness predictions are similar for rules that output opposite outcomes, we are estimating that contradictory rules have equivalent probabilities of being correct, which leads to lower confidence in the final outcome obtained combining those rules. 

Following that reasoning, a reliability estimation (probability) of the system for a patient $n$ can be theoretically obtained by comparing the means of the predicted rules’ correctness (PRC) for the two classes. More specifically:

\begin{equation}
    Reliability_{n}
    =  \left |\frac{1}{K_{n}^{+}} \sum_{j=1}^{K_{n}^{+}} (PRC)_{j,n} - \frac{1}{K_{n}^{-}}\sum_{l=1}^{K_{n}^{-}} (PRC)_{l,n} \right | 
    \label{eq:reliab}
\end{equation}

where $K_{n}^{+}$ is the set of rules that output the positive class (label=1), and $K_{n}^{-}$ is the set of rules that output the negative class (label=0), for the patient $n$.

For example, for the patient analyzed in Table \ref{tab:case_study}, the mean correctness for positive rules is 80.5\% and for negative rules is 48.0\%. Thus, the reliability of the decision support system for this patient is considerably low (32.5\%).

As the reliability estimation assesses the confidence in the predicted outcome, it is expected to attribute higher values to the patients that are well classified and lower values to the ones wrongly classified. In other words, lower reliability should be associated with a higher misclassifications rate.  

\begin{figure}[ht]
\vskip 0.2in
\begin{center}
\centerline{\includegraphics[width=\columnwidth]{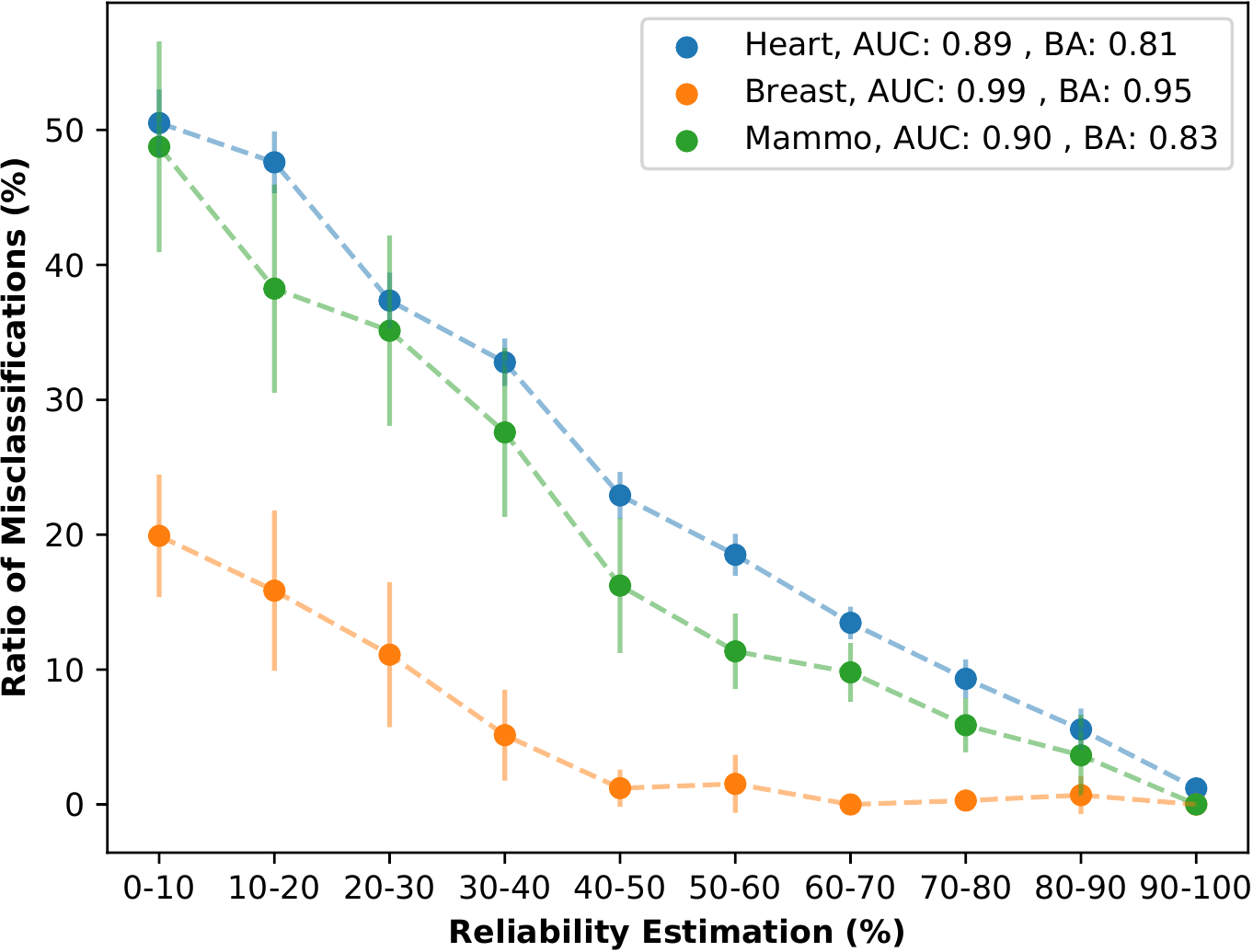}}
\caption{Variation of misclassifications rate depending on the reliability probabilities. The reliability curves are related to the mean and 95\% confidence interval obtained using the 5-times repeated  5-fold cross-validation.}
\label{fig:reliability}
\end{center}
\vskip -0.2in
\end{figure}

In order to inspect such, the decision system was applied to the three clinical datasets (Heart, Breast, Mammo) as described in section \ref{subsec:personalization}, and the reliability probabilities were estimated as detailed in this section. Figure \ref{fig:reliability} shows how the misclassifications rate varies with the predicted reliability estimation, for the three datasets. The corresponding ROC AUC and balanced accuracy (BA), related to the outcome predictions, are also presented to support the analysis.

As desired, Figure \ref{fig:reliability} shows that the reliability estimation presents a very noticeable relation with the misclassifications rate. More specifically, there are high rates of incorrect classifications when the CDSS is predicted to have low reliability (range 0-10\%), which decrease as the reliability increases. Moreover, for top reliability estimation values (range 90-100\%), it approaches the zero incorrect classifications.

Naturally, those curves depend on the overall predictive performances. The higher the ability to classify the patients, the lower the ratio of misclassifications. Particularly, for the Breast dataset, the model has very good discrimination (BA=95\%), and thus for reliability scores higher than 40\% there are very few misclassifications, which is not verified for the other scenarios.

The results visualized in Figure \ref{fig:reliability} suggest that the CDSS can provide a good estimation of the reliability of its own predictions. That information can be then used by the clinicians to know when they can trust the algorithm, or when it has a higher likelihood of providing incorrect decisions/recommendations. In short, it allows to better assess the system’s behave for a particular patient, providing a higher level of personalized interpretability.

\subsection{Some Final Considerations}

As summarized in Figure \ref{fig:cdss_scheme}, this novel clinical decision support system presents, for a given patient, (1) the output of each rule from the interpretable decision set, (2) the predicted probability of each rule being correct, (3) the computed outcome, and (4) an estimation of the CDSS's reliability. It gives valuable information to assist the decision-making in healthcare. Besides that, the CDSS has shown to achieve predictive performances equivalent to standard ML methods  \cite{Valente2021, valente2021improving}. While we presented and detailed how the several elements are combined to form a new CDSS (sections \ref{subsec:decision set}-\ref{subsec:reliability}), they can be individually optimized. More specifically, in each of them, different approaches may be followed: (a) the methodology to obtain the decision set, (b) the ML method used to train the prediction models, (c) the personalized voting technique, and eventually, (d) the formulation used to obtain the reliability estimation. All those individual procedures have a significant impact on the obtained CDSS, and thus their improvement leads to an improvement of the CDSS itself.

\section{Conclusion}

In this paper, we aimed at reviewing the general applicability of a recently developed clinical decision support system. Moreover, we intended to present and discuss how its several elements contribute to enhancing the CDSS's interpretability. We have described how a decision set can be applied to generate simultaneously global and local interpretability, while providing personalized predictions and reliability estimations. Furthermore, we have shown it has the potential to improve the predictive ability of the decision set. Such properties appear to be valuable contributions to healthcare scenarios. However, some considerations require further analysis and validation to improve its broader application. First, the use of a decision set composed of one-way decision rules. Rule-based methodologies often employ such rules, so it may promote the incorporation of literature approaches within the presented methodology. Second, the application to non-binary scenarios. The adaption to multiclass problems seems feasible, probably through a one-versus-all procedure. The application to regression problems looks more complicated and not direct, as the approach compares two discrete values to obtain the rule's correctness label, being its prediction a classification problem.

\section*{Acknowledgements}

This work was supported by Fundacão para a Ciência e Tecnologia (FCT), under the lookAfterRisk research project (POCI-01-0145-FEDER-030290).

\bibliography{references_icmlws}
\bibliographystyle{icml2021}

\end{document}